\documentclass[11pt,a4paper]{article}
\usepackage[left=15mm, top=10mm, right=15mm, bottom=20mm, nohead=true, nofoot=true]{geometry}

\usepackage[utf8]{inputenc}
\usepackage{authblk}
\usepackage{indentfirst}
\usepackage{misccorr}
\usepackage{graphicx}
\usepackage{amsmath}
\usepackage{amssymb}
\usepackage{multicol}
\usepackage{bm}
\usepackage{longtable}
\usepackage{pdflscape}
\usepackage{epstopdf}
\usepackage{multirow}
\usepackage{adjustbox}
\usepackage{amsfonts}
\usepackage{ifthen}
\usepackage{fancyhdr}
\usepackage{setspace}
\usepackage{xcolor}
\usepackage[round,authoryear,comma,sort&compress,numbers]{natbib}
\usepackage[toc,title,page]{appendix}
\usepackage[hidelinks]{hyperref}
\usepackage{url}
\usepackage{natbib}

\hypersetup{
    colorlinks=true,
    linkcolor=green,
    citecolor=blue,
    filecolor=magenta,
    urlcolor=cyan,
    pdftitle={Sharelatex Example},
    pdfpagemode=FullScreen,
}

\fancypagestyle{plain}{\chead{\texttt{\textcolor{brown}{Communications of BAO, Vol. ?, Issue ?, 20??, pp. ???-???}}}}
\title{\textbf{Detection of the Long Period
Variable Stars of And\,II Dwarf Satellite galaxy}}
\author[1, 2]{Hedieh Abdollahi \thanks{hediehabdollahi@ipm.ir, Corresponding author}}
\author[1]{Atefeh Javadi}

\author[3]{Jacco Th. van Loon}
\author[4, 5]{Iain McDonald}
\author[1]{Mahdi Abdollahi}
\author[6, 7, 8]{Elham Saremi}
\author[1, 9]{Habib G. Khosroshahi}
\author[1]{Hamidreza Mahani}

\affil[1]{\scriptsize School of Astronomy, Institute for Research in Fundamental Sciences (IPM), P.O. Box 1956836613, Tehran, Iran}
\affil[2]{\scriptsize Konkoly Observatory, HUN-REN Research Centre for Astronomy and Earth Sciences, MTA Centre of Excellence, Konkoly-Thege Mikl\'os \'ut 15-17, H-1121, Budapest, Hungary}

\affil[3]{\scriptsize Lennard-Jones Laboratories, Keele University, ST5 5BG, UK}
\affil[4]{\scriptsize Department of Physical Sciences, The Open University, Walton Hall, Milton Keynes, UK}
\affil[5]{\scriptsize Jodrell Bank Centre for Astrophysics, Alan Turing Building, University of Manchester, M13 9PL, UK}

\affil[6]{Instituto de Astrofísica de Canarias, Calle Vía L'actea s/n, E-38205 La Laguna, Spain}
\affil[7]{Departamento de Astrofísica, Universidad de La Laguna, E-38205 La Laguna, Spain}
\affil[8]{School of Physics \& Astronomy, University of Southampton, Highfield Campus, Southampton SO17 1BJ, UK}
\affil[9]{\scriptsize Iranian National Observatory, Institute for Research in Fundamental Sciences (IPM), Tehran, Iran}

\def\apj{Astrophys.~J. }

\def\pasp{Publ.~Astron.~Soc.~Pac. }

\def\aap{Astron.~Astrophys. }

\def\apjl{Astrophys.~J.~Lett. }

\def\aj{Astron.~J. }
\def\mnras{Mon.~Not.~R.~Astron.~Soc. }

\def\aapr{Astron.~Astrophys.~Rev. }

\begin{document}
\pagestyle{empty}
\newpage
\pagestyle{fancy}
\label{firstpage}
\date{}
\maketitle

\begin{abstract}
We conducted an extensive study of the spheroidal dwarf satellite galaxies around the Andromeda galaxy to produce an extensive catalog of LPV stars. The optical monitoring project consists of 55 dwarf galaxies and four globular clusters that are members of the Local Group. We have made observations of these galaxies using the WFC mounted on the 2.5 m INT in nine different periods, both in the $i$-band filter Sloan and in the filter $V$-band Harris. We aim to select AGB stars with brightness variations larger than 0.2 mag to investigate the evolutionary processes in these dwarf galaxies. The resulting catalog of LPV stars in Andromeda's satellite galaxies offers updated information on features like half-light radii, TRGB magnitudes, and distance moduli. This manuscript will review the results obtained for And\,II galaxy. Using the Sobel filter, we have calculated the distance modulus for this satellite galaxy, which ranges from 23.90 to 24.11 mag.
\end{abstract}
\emph{\textbf{Keywords:} stars: evolution --
stars: AGB and LPV--
stars: luminosity function, mass function --
stars: mass-loss --
stars: oscillations --
galaxies: stellar content
galaxies: Local Group}

\section{Introduction} \label{sec:intro}

Dwarf galaxies are known to show a large range of properties due to the complications within the galactic environment-affecting star formation rates, gas composition, and interactions with other galaxies \citep{2013pss6.book..207V}. Within the Local Group, they are noteworthy for their proximity, their variety, and the large range in metallicity and morphology. Their simpler star formation history compared to other galaxies makes them easier to study. The studies of dwarf galaxies have the potential to provide clues on galaxy formation scenarios, both the hierarchical and downsizing models. Dwarf galaxies are very crucial in attempts to understand dark matter since they are dominated by this form of matter, unlike globular star clusters. Their distribution in galaxy clusters may indicate the distribution of dark matter; hence, they may serve as better indicators of mass distribution within clusters than brighter galaxies. 

Generally, dwarf galaxies have low stellar density and have undergone minimal dynamical evolution, preserving their initial mass function. Several questions remain about galaxy evolution, such as the differences between the evolutionary processes of dwarf satellites and isolated dwarfs, and the impact of gas removal mechanisms on star formation. Focusing on dwarf galaxies, particularly Andromeda satellites, the Isaac Newton telescope monitoring survey of dwarf galaxies in the Local Group (INT survey) was conducted to study their evolutionary history. Observations were made at three-month intervals from June 2015 to February 2018. To study galaxy evolution, asymptotic giant branch (AGB) stars were observed as they are key tracers of SFH \citep{2019MNRAS.483.4751H, 2017MNRAS.466.1764H, 2017MNRAS.464.2103J, 2014MNRAS.445.2214R, 2011ASPC..445..497J}. AGB stars, which can become long-period variable (LPV) stars, provide insights into the dynamics and evolution of dwarf galaxies \citep{2016MmSAI..87..278J, 2013MNRAS.432.2824J}. Stars with initial masses between 0.5 to 8 M$_\odot$ can reach the late stages of the AGB evolutionary path \citep{2018A&ARv..26....1H}. AGB stars reach their maximum luminosity during these stages, making them easier to observe in faint dwarf galaxies. The luminosity of LPVs is linked to their initial birth mass, providing clues about their mass, age, and pulsation duration. AGB stars are excellent indicators for reconstructing SFH as they age from 10 Myr to 10 Gyr \citep{2011MNRAS.414.3394J}.

Table \ref{table:objects} lists the M\,31 satellites in this study, detailing the observational time, epoch, filter, exposure time, airmass, and seeing conditions for each galaxy. Of the 24 dwarf galaxies in the Andromeda system, four (M\,110, M\,32, Pisces I, and Pegasus) lacked sufficient observational epochs to detect long-period variable (LPV) stars, while three had already been studied previously \citep{2023ApJ...948...63A,2021ApJ...923..164S,2021ApJ...910..127N}.

\begin{figure}[ht]
\centering
\includegraphics[width=0.6\textwidth]{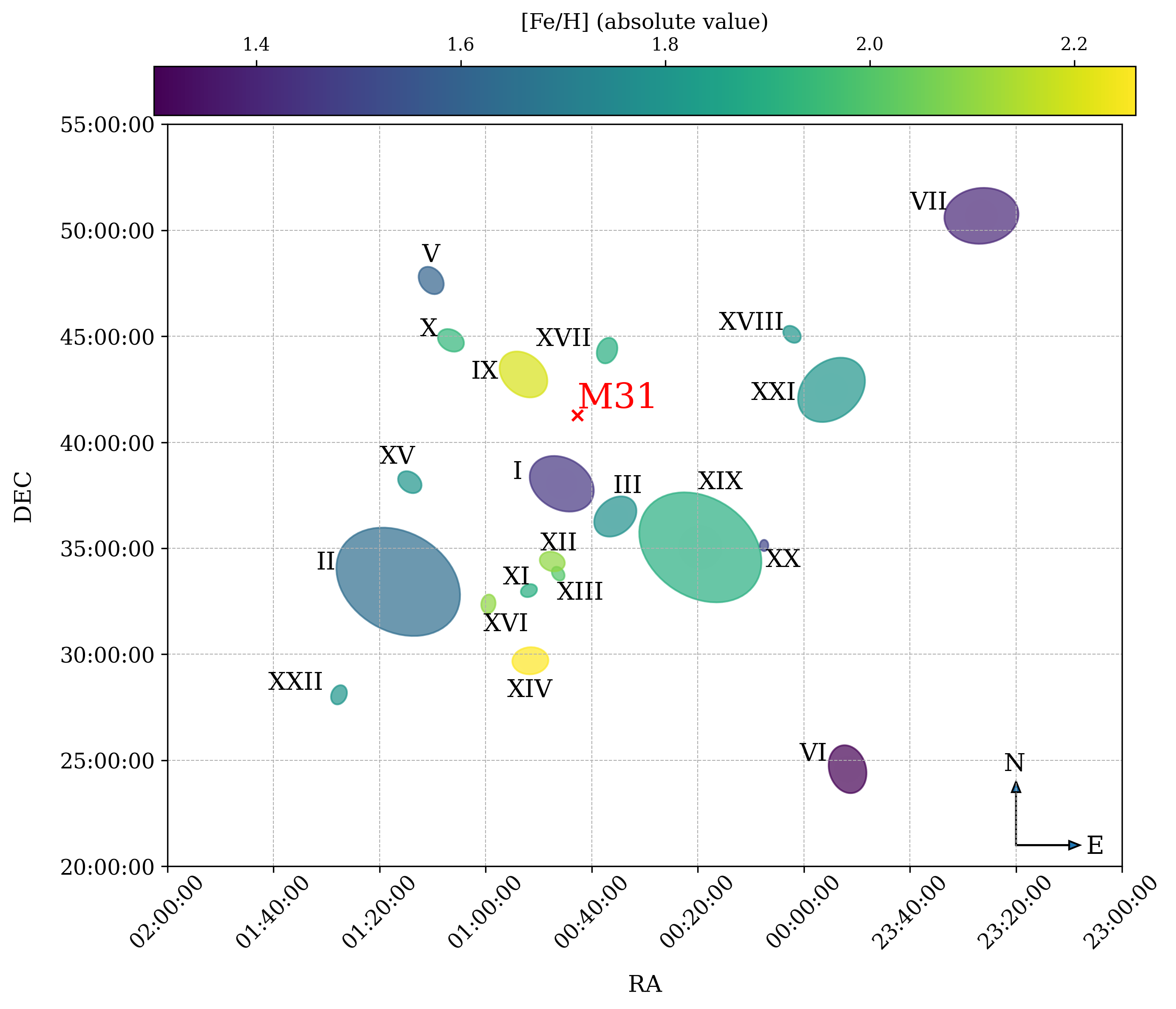}
\caption{Satellite galaxies of Andromeda. The semi-major axis of the galaxies is represented proportionally to their actual sizes, based on data from Table \ref{table:objects}.}
\label{fig:M31_map}
\end{figure}

As illustrated in Figure \ref{fig:M31_map}, And\,II is significantly larger than the other Andromeda satellites, except for And XIX. And\,II was first identified through the analysis of images acquired with the 48-inch Schmidt telescope at Palomar Observatory, during observations conducted by Sidney Van den Bergh in 1970 and 1971 \citep{1972ApJ...171L..31V}. In this paper, we present the results of time-series observations of this galaxy, performed using the INT Telescope.

\begin{table*}
\caption{Observational properties and half-light radius of targets.}
\small 
\setlength{\tabcolsep}{1pt}
\centering
\begin{tabular}{lllllllllll}
\hline\hline

 \noalign{\smallskip}
{Galaxy}           &
{R.A. $^a$}             &
{Dec $^a$}              &
{$\epsilon^b$} & 
{[Fe/H]$^c$}        &
{$r_{\rm h}$ $^b$}   &
{$r_{\rm h}$ $^{Exponential}$ }    &
{$r_{\rm h}$ $^{Plummer}$ }        &
{$r_{\rm h}$ $^{S\acute{e}rsic}$ } &
N$_{Total}$ &
N$_{LPV}$\\

 &
(J2000) &
(J2000) &
 & 
(dex)   &
(arcmin)   &
(arcmin)   &
(arcmin)     &
(arcmin)   &&\\

\hline
\multicolumn{11}{l}{}\\ 
And\,I $^d$ & 00 45 39.80 & $+38$ 02 28.00 & $0.28\pm0.03$ & $-1.45\pm0.04$ & $3.90\pm0.10$ & $3.20\pm0.30$ & - & - & 10243 & 470\\

And\,II  & 01 16 29.78 & $+33$ 25 08.75 & $0.16\pm0.02$ & $-1.64\pm0.04$ & $5.30\pm0.10$ & 5.32$^{+0.19}_{-0.02}$ & 5.21$^{+0.09}_{-0.12}$ & 5.35$^{+0.16}_{-0.05}$ & 10384 & 728\\

And\,III & 00 35 33.78 & $+36$ 29 51.91 & $0.59\pm0.04$ & $-1.78\pm0.04$ & $2.20\pm0.20$ &  2.33$^{+0.07}_{-0.20}$ & 2.21$^{+0.19}_{-0.08}$ & 2.15$^{+0.25}_{-0.02}$ & 5409  & 573 \\

And\,V & 01 10 17.10 & $+47$ 37 41.00 & 0.26$^{+0.09}_{-0.07}$ & $-1.6\pm0.3$  & 1.60$^{+0.20}_{-0.10}$ &  1.88$^{+0.04}_{-0.17}$ & 1.84$^{+0.08}_{-0.14}$ & 1.86$^{+0.06}_{-0.16}$ & 7928 & 398\\

And\,VI & 23 51 46.30 & $+24$ 34 57.00 & $0.41\pm0.03$ & $-1.3 \pm0.14$ & $2.30\pm0.20$ & 2.24$^{+0.15}_{-0.03}$ & 2.00$^{+0.02}_{-0.16}$ & 2.27$^{+0.12}_{-0.07}$ & 6336 & 364\\

And\,VII $^e$ & 23 26 31.74 & $+50$ 40 32.57 & $0.13\pm0.04$ & $-1.40\pm0.30$ & $3.50\pm0.10$ & $3.80\pm0.30$ & - & - & ? & 55\\

And\,IX $^f$ & 00 52 53.00 & $+43$ 11 45.00 & 0.00$^{+0.16}_{-0.00}$ & $-2.2\pm0.2$ & 2.00$^{+0.30}_{-0.20}$ & $2.50\pm0.26$ & - & - & 8653 & 77\\

And\,X & 01 06 33.70 & $+44$ 48 15.80 & 0.10$^{+0.34}_{-0.10}$ & $-1.93\pm0.11$ & 1.10$^{+0.40}_{-0.20}$ & 1.30$^{+0.35}_{-0.20}$ & 1.29$^{+0.36}_{-0.19}$  & 1.28$^{+0.37}_{-0.18}$ & 4025 & 418\\

And\,XI & 00 46 20.00 & $+33$ 48 05.00 & 0.19$^{+0.28}_{-0.19}$  & $-2.0\pm0.2$ & $0.60\pm0.20$ &  0.54$^{+0.18}_{-0.06}$ & 0.53$^{+0.19}_{-0.05}$ &  0.53$^{+0.19}_{-0.05}$ & 3999 & 495 \\

And\,XII & 00 47 27.00 & $+34$ 22 29.00 & 0.61$^{+0.16}_{-0.48}$  & $-2.1 \pm0.2 $ & 1.80$^{+1.20}_{-0.70}$ & 1.72$^{+0.33}_{-0.18}$ & 1.72$^{+0.33}_{-0.18}$ &  1.73$^{+0.33}_{-0.18}$ & 3555 & 237\\

And\,XIII & 00 51 51.00 & $+33$ 00 16.00 & 0.61$^{+0.14}_{-20}$ & $-1.9\pm0.2$ & 0.80$^{+0.40}_{-0.30}$ & 0.71$^{+0.25}_{-0.07}$ & 0.69$^{+0.27}_{-0.05}$ & 0.73$^{+0.23}_{-0.09}$ & 3794 & 587\\

And\,XIV & 00 51 35.00 & $+29$ 41 49.00 & 0.17$^{+0.16}_{-0.17}$ & $-2.26\pm0.05$ & $1.50\pm0.20$ & 1.67$^{+0.20}_{-0.17}$ & 1.66$^{+0.21}_{-0.16}$   & 1.66$^{+0.21}_{-0.16}$ & 7876 & 360\\

And\,XV & 01 14 18.70 & $+38$ 07 03.00 & $0.24\pm0.10$ & $-1.8\pm0.2$ & $1.30\pm0.10$ & 1.55$^{+0.01}_{-0.25}$ & 1.48$^{+0.08}_{-0.18}$ & 1.46$^{+0.10}_{-0.16}$ & 7215 & 439 \\

And\,XVI & 00 59 29.80 & $+32$ 22 36.00 & $0.29\pm0.08$ & $-2.1\pm0.2$ & $1.00\pm0.10$ & 1.19$^{+0.01}_{-0.19}$ & 1.14$^{+0.06}_{-0.14}$ & 1.13$^{+0.07}_{-0.13}$ & 2972 & 183 \\

And\,XVII  & 00 37 07.00 & $+44$ 19 20.00 & $0.50\pm0.10$ & $-1.9\pm0.2$ & $1.48\pm0.30$ & 1.48$^{+0.01}_{-0.17}$ & 1.47$^{+0.02}_{-0.17}$ & 1.48$^{+0.02}_{-0.17}$ & 10369 & 430 \\

And\,XVIII & 00 02 14.50 & $+45$ 05 20.00 & 0.03$^{+0.28}_{-0.03}$ & $-1.8\pm0.1$ & $0.80\pm0.10$ & 0.97$^{+0.15}_{-0.01}$ & 0.95$^{+0.01}_{-0.15}$ & 0.93$^{+0.03}_{-0.13}$ & 2638 & 266\\

And\,XIX & 00 19 32.10 & $+35$ 02 37.10 & 0.58$^{+0.05}_{-0.10}$ & $-1.9\pm0.1$ & 14.20$^{+3.40}_{-1.90}$ & 14.28$^{+0.49}_{-0.08}$ & 14.22$^{+0.55}_{-0.02}$ & 14.44$^{+0.33}_{-0.24}$ & 7441 & 1676 \\

And\,XX & 00 07 30.70 & $+35$ 07 56.40 & 0.11$^{+0.41}_{-0.11}$ & $-1.5\pm0.1$ & 0.40$^{+0.20}_{-0.10}$ & 0.50$^{+0.04}_{-0.10}$ & 0.49$^{+0.05}_{-0.09}$ & 0.46$^{+0.07}_{-0.06}$ & 4050 & 654\\

And\,XXI & 23 54 47.70 & $+42$ 28 15.00 & 0.36$^{+0.10}_{-0.13}$ & $-1.8\pm0.2$ & 4.10$^{+0.80}_{-0.40}$ & 3.83$^{+0.27}_{-0.55}$ & 3.82$^{+0.28}_{-0.54}$ & 3.82$^{+0.28}_{-0.54}$ & 2621 & 293\\

And\,XXII & 01 27 40.00 & $+28$ 05 25.00 & 0.61$^{+0.10}_{-0.14}$ & $-1.8$ & 0.90$^{+0.30}_{-0.20}$ & 0.90$\pm0.18$ & 0.87$^{+0.21}_{-0.15}$ &  0.85$^{+0.23}_{-0.13}$ & 4645 & 295\\

\hline
\end{tabular}
\footnotesize{$^a$ Coordinates inferred from the \cite{NED_Full} portal.}\\
\footnotesize{$^b$ All ellipticities are referred from \cite{2016ApJ...833..167M}, except for And\,VI which is from \cite{2012AJ....144....4M}.}\\
\footnotesize{$^c$ \cite{2012AJ....144....4M}, $^d$ \cite{2020ApJ...894..135S} , $^e$ \cite{2021ApJ...910..127N}, and $^f$ \cite{2023ApJ...948...63A}. }\\
\footnotesize{$\epsilon$ = $1-b/a$, where $b$ is the semi-minor axis and $a$ is the semi-major axis.}
\label{table:objects}
\end{table*}

\section{Observation and Data} \label{sec:data}

The Wide Field Camera (WFC) on the 2.5 m INT telescope at the Observatorio del Roque de los Muchachos in La Palma is equipped with four 2048$\times$4096 CCDs, each with a pixel scale of 0.33 arcseconds.

Long-period variable stars exhibit distinctive brightness fluctuations due to intrinsic mechanisms such as pulsations and eclipses. Time-series observations enable the quantification of periodicity, the detection of transient events, and the study of different evolutionary phases. This approach reveals detailed behaviors such as light-curves, amplitude modulation, and phase shifts. To investigate the magnitude variations of the LPV in And\,II, time series observations were performed using Sloan $i$, Harris $V$, and RGO $I$ filters with the WFC to analyze photometric variability.
The $i$-band filter enhances the contrast of spectral energy distributions (SEDs) for cool, evolved stars, particularly LPVs. The $V$ filter, with a peak transmission around 528 nm, is used to determine stellar color, which is directly related to temperature.

\section{Methodology} \label{sec:method}

\subsection{Data reduction and Photometry}

To obtain the science images of And\,II from raw data, the images were processed using the 'THELI' pipeline, which is optimized for precise astrometry and multi-chip CCD cameras.

Photometry in both $i$ and $V$ filters was performed using the {\sc \texttt{DAOPHOT/ALLSTAR}} package \citep{1987PASP...99..191S}. Initially, a selection of approximately 30-40 isolated stars located at various positions in the field was made using the {\sc \texttt{PSF}} routine. The purpose was to construct a point-spread function (PSF) model for each image following the {\sc \texttt{FIND}} and {\sc \texttt{PHOT}} procedures. A master image was then created by combining individual images through the {\sc \texttt{DAOMATCH}}, {\sc \texttt{DAOMASTER}}, and {\sc \texttt{MONTAGE2}} routines. This master image was used to generate a star list with the {\sc \texttt{ALLSTAR}} routine. Subsequently, the {\sc \texttt{ALLFRAME}} routine used the star list to estimate the instrumental magnitudes of stars by fitting the PSF models to the individual images \citep{Stetson94}.
The transformation of instrumental magnitudes into the standard system was accomplished using observations of standard stars \citep{Landolt92} and the {\sc \texttt{NEWTRIAL}} routine \citep{Stetson96}.

\subsection{Calibration and photometry assessment}

The photometric calibration process was conducted in three stages. First, aperture corrections were applied using the {\sc \texttt{DAOGROW}} and {\sc \texttt{COLLECT}} routines to calculate the differences in magnitude between the PSF-fitting and largest aperture photometry of about 40 isolated bright stars in each frame \citep{Stetson90}. The {\sc \texttt{NEWTRIAL}} routine then adjusted these aperture corrections for all stars in each frame. 

Second, the transformation to the standard photometric system was carried out by constructing transformation equations for each frame, which accounted for zero-point and atmospheric extinction. The mean of other zero-points was used for frames lacking a standard-field observation. The {\sc \texttt{CCDAVE}} routine applied these transformation equations to the program stars for each frame, and the {\sc \texttt{NEWTRIAL}} routine subsequently calibrated all other stars using the program stars as local standards. 

Finally, relative photometry between epochs was conducted to accurately distinguish variable from non-variable sources. Approximately 1000 common stars were selected across all frames within the magnitude interval 18 to 21 mag. Each star's deviation at each epoch was determined relative to the mean magnitude calculated from all epochs. These mean magnitudes were computed by weighting the individual measurements. The resulting corrections were then applied to the frames.

To evaluate the completeness of the survey, artificial stars were added using the {\sc \texttt{ADDSTAR}} routine \citep{1987PASP...99..191S} in both $i$- and $V$-band single frames, across discrete 0.5 mag bins ranging from 16 to 24.5 mag. The fraction of recovered artificial stars was estimated with the {\sc \texttt{ALLFRAME}} routine. The results indicated that the survey is sufficiently complete up to 22 mag in the $i$ and $V$ bands, near the tip of the RGB, and up to 50\% complete for stars with a magnitude of approximately 23 mag in both filters, confirming that nearly the entire AGB and RSG populations are detected for the primary research purpose.

\subsection{Detection of long-period variables}

The foreground stars were removed by cross-correlating the catalog with the Gaia DR3 catalog before LPV cognition \citep{GaiaDR3}.
The method developed by \citet{Welch93} and refined by \citet{Stetson96} involves calculating the Stetson variability index. This index quantifies a star’s brightness variability based on time series observations, considering the measured magnitudes and their errors.
The next step involves calculating the pairwise product of these standardized deviations. For variable stars, it is expected that the magnitudes in the $i$ and $V$ filters will either increase or decrease together across different observational nights within a time interval of less than half a minimum period (60 days for LPVs \citep{McDonald16}). Although the magnitude of change could vary, for each observation pair, the direction of change should remain the same to yield a positive product that measures the coherence of the deviations: if the deviations from the mean are consistent and correlated across observations, it will be positive to indicate variability; otherwise, random noise cancels out due to uncorrelated deviations.
The index of variability, $L$, normalizes and weights them based on the number of observations. A large value of $L$ denotes that the large probability of a star being an LPV is hence very useful for selecting the variable stars out of an extensive dataset.

The accuracy of the variability indices method is confirmed through simulations, such as the {\sc \texttt{ADDSTAR}} subroutine. This includes introducing artificial stars with known properties into the data and evaluating how effectively the method retrieves these inputs. This procedure aids in gauging the method's dependability and accuracy.

\subsection{Amplitude of Variability for Candidate Stars}

The variability amplitude was estimated by assuming a sinusoidal light-curve shape. By comparing the standard deviation in the magnitudes to the value expected for a completely random sampling of a sinusoidal variation (0.701), the amplitude can be determined using the equation that defines it as the difference between the minimum and maximum brightness.

In this study, we prioritize stars with amplitudes exceeding 0.2 mag due to our uncertainty regarding the nature of stars with lower amplitudes.

\begin{equation}
A=2\sigma/0.701,
\label{eq:Amp_i}
\end{equation}

\section{Results} \label{sec:results}

The photometric analysis of the And\,II dwarf galaxy has not only led to the discovery of LPV stars but has also aided in the estimation of the tip of the red giant branch (TRGB). This estimation is crucial for calculating the distance modulus, which is fundamental for comprehending the spatial arrangement of galaxies. Moreover, the analysis offers a method for determining the half-light radius of these galaxies.

\begin{figure}[ht]
\centering
  \includegraphics [width=0.4\textwidth]{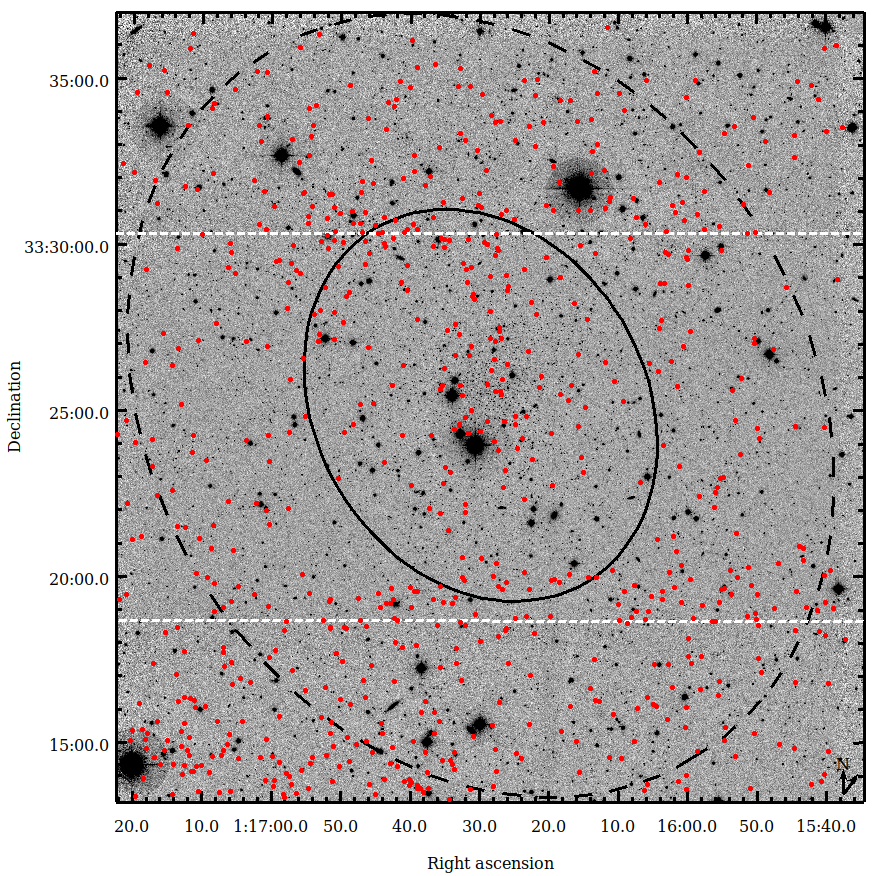}
  \includegraphics [width=0.4\textwidth]{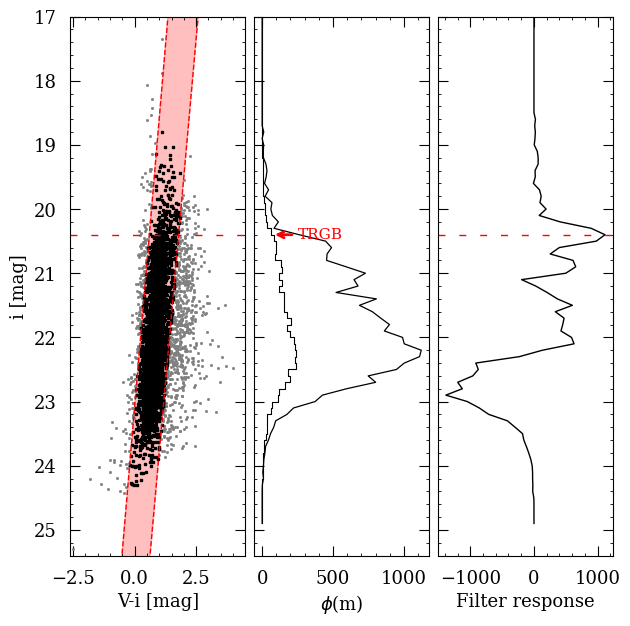}
  \includegraphics [width=0.4\textwidth]{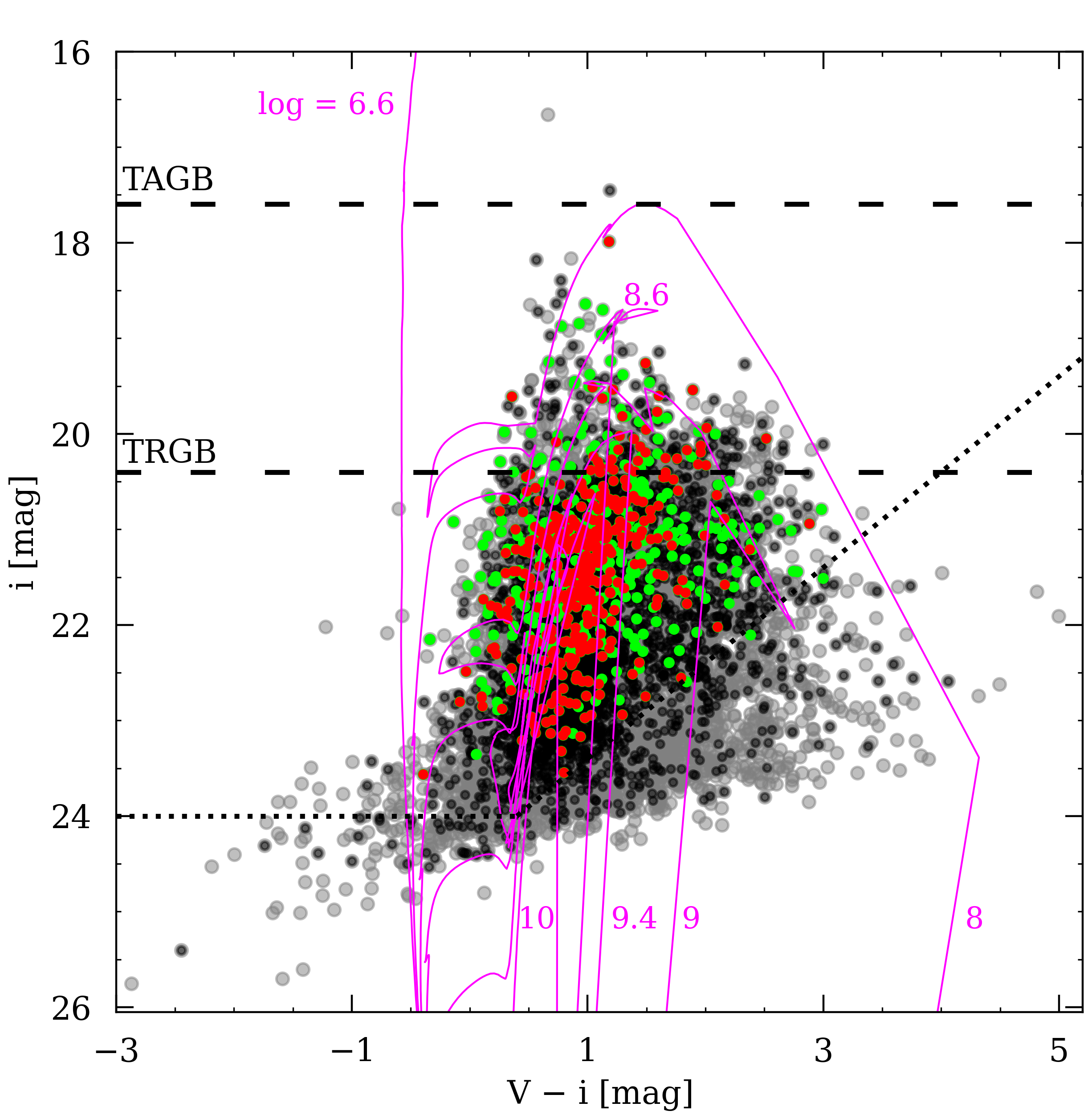}
  \includegraphics [width=0.4\textwidth]{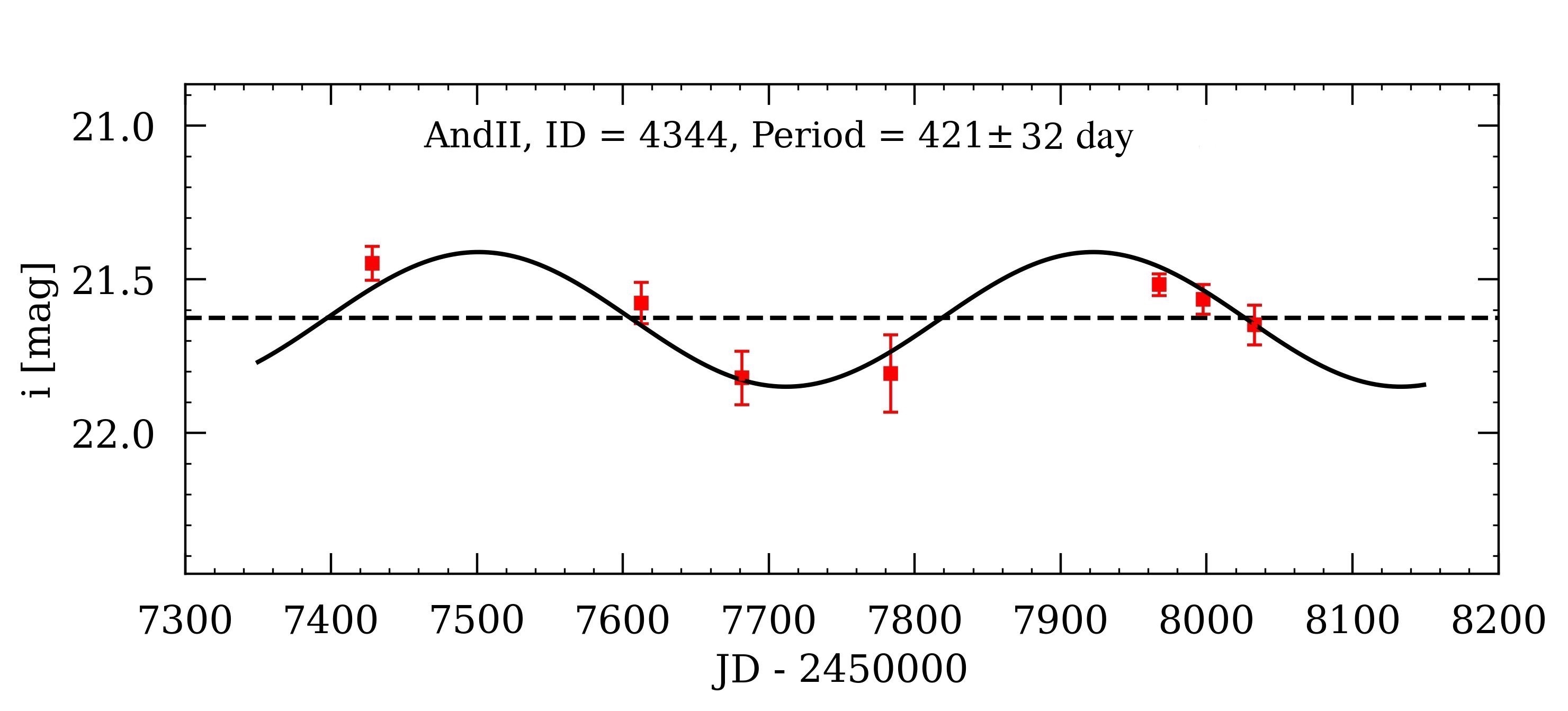}
  \caption{Distribution of the LPVs, detection of the TRGB, CMD, and light-curve in the And\,II galaxy.}.
\label{fig:result}  
\end{figure}

In Figure \ref{fig:result}, the upper left subplot illustrates the distribution of LPV candidates in the studied areas. Red circles indicate LPV candidates, while solid and dashed black circles represent the half-light radius and twice the half-light radius of the dwarf galaxy, respectively. The black arrows point toward the center of the Andromeda galaxy. As illustrated in this subplot, photometric measurements required the utilization of CCD1, CCD3, and CCD4 of WFC due to the considerable size of this galaxy. As a result, the total number of stars identified (10, 384) and the detected LPV stars (825) are notably higher than those found in other target galaxies.

In the upper right subplot, the left panel shows stellar sources within two half-light radii, the middle panel presents the histogram of the luminosity function, and the right panel displays the Sobel filter response for edge detection of the tip of the red giant branch (TRGB). These plots are specific to And\,II and resulted in TRGB = 20.40$\pm$0.10 mag (The TRGB is highlighted with red lines and arrows). The distance modulus computed for this galaxy stands at 23.81$\pm$0.10 mag ($\sim$ 578 kpc). The results for the TRGB and distance modulus align with those reported in the study by \citet{McConnachie04}.

The lower left subplot shows the color-magnitude diagram (CMD) of the And\,II dwarf satellite galaxy.
The gray points denote stars that were identified photometrically in the final images. The black dots represent stars situated within two times the half-light radius of the center of the target galaxy. Green points indicate potential long-period variable candidates throughout the entire studied field, while red points highlight LPV candidates within two times the half-light radius. The magenta lines illustrate the relevant isochrones.

Finally, the lower right subplot displays the light-curve of a sample variable star in And\,II galaxy.

\section{Summary} \label{sec:Summary}

We utilized the Wide Field Camera (WFC) on the Isaac Newton Telescope (INT) to conduct observations aimed at monitoring most of the dwarf galaxies observable from the northern hemisphere. Our observations primarily focused on the $i$-band filter, with additional $V$-band observations carried out over up to nine epochs. This study presents the initial findings for galaxy And\,II, showcasing our methodology and the potential scientific insights that this project can provide.

For each of these galaxies, we developed photometric catalogs focused mainly on the area covered by CCD4 of the WFC, which spans $11.26^{\prime\prime} \times 22.55^{\prime\prime}$. These catalogs are comprehensive, providing both extensive photometric data and the identification of potential long-period variable (LPV) stars, especially those exhibiting amplitude variations greater than 0.2 mag. We derived the distance modulus for these galaxies by analyzing the tip of the RGB in the photometric data. Additionally, we measured the half-light radii for these satellite galaxies.

In future papers in this series, we will utilize these catalogs to explore the star formation history and dust production of all identified LPV candidates across the monitored galaxies. Furthermore, we investigate how color, and consequently temperature, changes during variability phases. By analyzing these changes and luminosity data, we aim to determine the variations in stellar radius and their correlation with mid-infrared excess.

\section*{\small Acknowledgements}
\scriptsize{We sincerely thank BAO for their warm hospitality during the conference, which greatly enhanced our experience and contributed to our success. We also appreciate the support from the H.A. via the ‘SeismoLab’ KKP-137523 Élvonal grant from the Hungarian Research, Development and Innovation Office (NKFIH).}

\scriptsize
\bibliographystyle{ComBAO}
\nocite{*}
\bibliography{ComBAO}

\end{document}